\documentclass[aps,prc,groupedaddress,showpacs,manuscript]{revtex4-1}
\usepackage{graphicx}

\usepackage{colordvi}

\begin{document}

\title{Effect of the spin-orbit interaction on flows in heavy-ion collisions at intermediate energies}

\author {Chenchen Guo$\, ^{1,2,3}$,
Yongjia Wang$\, ^{3}$,
Qingfeng Li$\, ^{3}$\footnote{liqf@hutc.zj.cn},
and
Feng-Shou Zhang$\, ^{1,2,4}$\footnote{Corresponding author: fszhang@bnu.edu.cn}}

\affiliation{
1) Key Laboratory of Beam Technology and Material Modification of Ministry of Education, College of Nuclear Science and Technology,
Beijing Normal University, Beijing 100875, China \\
2) Beijing Radiation Center, Beijing 100875, China \\
3) School of Science, Huzhou University, Huzhou 313000, China\\
4) Center of Theoretical Nuclear Physics, National Laboratory of Heavy Ion Accelerator of Lanzhou, Lanzhou 730000, China\\
\\
 }
\date{\today}

\begin{abstract}
The effect of the spin-orbit coupling in heavy ion collisions is
investigated based on an updated version of the ultra-relativistic quantum
molecular dynamics (UrQMD) model, in which the Skyrme
potential energy density functional is employed. And in special, the spin-orbit coupling
effects on the directed and elliptic flows of free nucleons emitted
from $^{197}$Au+$^{197}$Au collisions as functions of both the beam
energy and the impact parameter are studied. Our results show that
the net contribution of the spin-orbit term to flows of nucleons is
negligible, whereas a directed flow splitting between spin-up
and spin-down nucleons is visible especially at large impact parameters and a peak of the splitting is found at the beam energy
around 150 MeV$/$nucleon. We also found that the directed flow splitting between spin-up and spin-down neutrons is comparable with the neutron directed flow difference calculated by a soft and a stiff symmetry energy, indicating that the directed flow of neutrons cannot be used to pin down the stiffness of symmetry energy any more without considering the spin degree of freedom in models in case of spin polarization.

 \end{abstract}


\pacs{25.70.-z,24.10.-i,25.75.Ld}

\maketitle
\section{introduction}
It is well-known that the spin-orbit coupling plays important roles
in quantum systems such as atom and nucleus and so on. In the year
1949, Mayer and Jensen successfully interpreted the magic numbers by
introducing the spin-orbit coupling so that the shell structure of
nuclei can be clearly demonstrated~\cite{KU}. Afterwards, many
theoretical works related to the spin-orbit
coupling effects on properties of nuclei have been performed.

Nowadays, with the development of the so-called rare-isotope beam
facilities, much attention is being paid on the properties of
exotic nuclei \cite{Sorlin:2008jg}. Some
studies in recent decades have shown that magic numbers may change
when moving from stable to exotic nuclei \cite{Sorlin:2008jg,Sorlin:2012xr,Otsuka:2013hra}. A reduction in the
spin-orbit interaction was suggested to explain the appearance and
disappearance of the magic number in exotic
nuclei \cite{Schiffer:2004zz,nature}. Later on, the tensor force was further taken into account in a more complete picture \cite{Otsuka:2005zz}.
Therefore, the knowledge of density- and isospin- dependence of the
spin-orbit interaction (in-medium spin-orbit interaction) is
required. Heavy-ion collisions (HICs), being one of the
indispensable candidates for studying the nucleon-nucleon
interactions in nuclear medium especially away from normal
densities, ought to give support to this topic. However, although the
effect of spin-orbit interaction in the nuclear structure has been
extensively studied, its effect in the nuclear reaction is still poorly
known because of its weak contribution with normal vision.

Recently,
based on the time-dependent Hartree-Fock (TDHF) model, the effect of
spin-orbit interaction and tensor force in low-energy nuclear
reactions is studied~\cite{Umar:1986qs,Iwata:2010ws,Maruhn:2006uh}.
By incorporating the spin degree of freedom into the
isospin-dependent Boltzmann-Uehing-Uhlenbeck (IBUU) transport model,
the effect of spin-orbit interaction in HICs at intermediate
energies was firstly studied by Xu \emph{et al.}~\cite{Xu:2012hh,xu:2014}.
It was found that the nucleon spin up-down differential
transverse flow is sensitive to the spin-orbit interaction. In view
of the model dependence  in the description of HICs, it is
necessary to study the spin-orbit coupling effect on various
observables with other transport models. Especially, as sensitive
probes to the density dependence of symmetry energy, collective
flows of light clusters should be checked with care since the
contribution of symmetry energy to flows is also secondary when
comparing to the iso-scalar bulk terms
\cite{Cozma:2013sja,Wang:2013wca}.

The paper is arranged as follows. With the help of the Skyrme
potential energy density functional, the spin-orbit term is
introduced into the ultra-relativistic quantum molecular dynamics (UrQMD) model, which is shown in the next section. In Sec.
III, results of the spin-orbit coupling effect on directed and elliptic flows of free nucleons emitted from Au+Au collisions
with various impact parameters at intermediate energies are depicted and discussed.
 Finally, a summary is given in Sec. IV.

\section{Model description}
It is known that the UrQMD model \cite{Bass:1998ca,Bleicher:1999xi}
inherits analogous principles as the quantum molecular dynamics
(QMD) model \cite{aichelin91} in its mean-field part and the
relativistic quantum molecular dynamics (RQMD) model \cite{rQMD} in the
corresponding two-body collision part. It is successfully extended to describe HICs with beam
energy starting from as low as several tens of MeV/per nucleon (low
SIS) up to the highest one available at CERN Large Hadron Collider
(LHC) \cite{Li:2011zzp,Petersen:2006vm,Li:2012ta}.

In order to better and more systematically describe the experimental
data existing at SIS energies and especially to extract more confirmed
information of the density dependent nuclear symmetry energy,
the potential terms in UrQMD have been replaced globally with
the Skyrme potential energy density functional \cite{Wang:2013wca}. It is known that for
the nuclear interaction, the Hamiltonian $H$ consists of the kinetic
energy $T$ and the effective interaction potential energy $U$,

\begin{equation}
H=T+U,    \label{ham1}
\end{equation}
where $U$ includes the Coulomb $U_{Cou}$ and the Skyrme $U_{Sky}$
terms. The $U_{Sky}$ can be written as
~\cite{Vautherin:1971aw,Engel:1975zz}

\begin{equation}
U_{Sky}=\int u_{\rho,md,so} d\vec{r}. \label{Sky1}
\end{equation}
In Eq.~(\ref{Sky1}), the density- and momentum- dependent $u_\rho$ and
$u_{md}$ terms had been successfully incorporated into UrQMD
\cite{Wang:2013wca} while at present the energy density arising from
the spin-orbit interaction is further considered which consists of
time-even
~\cite{Vautherin:1971aw,Engel:1975zz,Maruhn:2006uh,Bender:2003jk,Maruhn:2013mpa}
\begin{eqnarray}
\label{eq2}
u_{so}^{even}=-\frac{1}{2}W_0(\rho \nabla\cdot\vec{J}+\rho_n \nabla\cdot\vec{J}_n + \rho_p \nabla\cdot\vec{J}_p),
\label{soeven}
\end{eqnarray}
and time-odd
\begin{eqnarray}
\label{eq3} u_{so}^{odd}=-\frac{1}{2}W_0[\vec{s}\cdot (\nabla\times
\vec{j}) +\vec{s}_n\cdot (\nabla\times \vec{j}_n )+\vec{s}_p\cdot
(\nabla\times \vec{j}_p)]
\end{eqnarray}
terms. In Eqs.~(\ref{eq2}) and (\ref{eq3}) the number density $\rho$,
spin density $\vec{s}$, momentum density $\vec{j}$ and spin-current
density $\vec{J}$ are given by
\begin{eqnarray}
&&\rho(\vec{r})=\sum_i\rho_i(\vec{r})=\sum_i \frac{1}{(2\pi
L)^{3/2}}e^{[-(\vec{r}-\vec{r}_i)^2/(2L)]},\\
&&\vec{s}(\vec{r})=\sum_i\rho_i(\vec{r})\vec{\sigma}_i,\\
&&\vec{j}(\vec{r})=\sum_i\rho_i(\vec{r})\vec{p}_i,\\
&&\vec{J}(\vec{r})=\sum_i\rho_i(\vec{r})\vec{p}_i\times\vec{\sigma}_i, \label{Jr}
\end{eqnarray}
where $L$ is the width parameter of the nucleon's wave packet and
set to be 2 $fm^{2}$ for Au+Au collisions. $\vec{r}_i$, $\vec{p}_i$ and
$\vec{\sigma}_i$ are the coordinate, momentum and spin of the
\emph{i}th nucleon, respectively.
If we ignore the isospin asymmetry and discontinuity in the nuclear medium, we can deduce from Eqs.~(\ref{soeven}-\ref{Jr}) that the spin-orbit potential is mainly controlled by the form $\nabla\rho\cdot(\vec{p}\times\vec{\sigma})$ of the nucleon, and which has opposite signs for spin-up and spin-down nucleons. Note, in the simulation of HICs, the reaction plane is usually defined with the $x$ (along the impact parameter vector) and $z$ (along the beam direction) axes. Therefore, the \emph{i}th nucleon is set to be spin-up (spin-down) if the $\vec{\sigma}_i$ in its \emph{y} direction  is positive (negative).
And, based on the current setting, the major effect comes from the $(\nabla\rho)_xp_z\sigma_y$ part since $p_z$ and $(\nabla\rho)_x$ are larger than the separate quantities in the other two directions.
In this work, in view of the present constraint on the incompressibility ($K_0\simeq200-260$ MeV)~\cite{Dutra:2012mb}, the SkP interaction
is chosen where the incompressibility $K_0=201$ MeV and the strength
of the spin-orbit coupling $W_0=0.1$ GeV fm$^5$.
The in-medium nucleon-nucleon cross section and Pauli blocking treatments in the collision term
are taken as the same way as our previous work in Ref.~\cite{Wang:2013wca}, which has described the recent FOPI flow data fairly well.

\section{Results and discussion}
The directed and elliptic flows have been commonly used as
experimental observables to probe the stiffness of nuclear equation of state
\cite{Danie02}. They can be quantified from the Fourier expansion of
the particle azimuthal anisotropic distributions with respect to the
reaction plane \cite{reisdorf1997},
 \begin{equation}
 \frac{dN}{d\phi} = v_0 [1 + 2v_1 \cos(\phi) + 2v_2
\cos(2\phi)] ,
\end{equation}
where $\phi$ is the azimuthal angle. The coefficient $v_1$ and $v_2$
represent the values of directed and elliptic flows, respectively,
and can be expressed as
 \begin{equation}
 v_1\equiv \langle
 cos(\phi)\rangle=\langle\frac{p_x}{p_t}\rangle,
 \label{eqv1}
 \end{equation}
 and
 \begin{equation}
 v_2\equiv \langle
 cos(2\phi)\rangle=\langle\frac{p_x^2-p_y^2}{p_t^2}\rangle.
 \label{eqv2}
 \end{equation}
Here $p_x$ and $p_y$ are the two components of the transverse
momentum $p_t=\sqrt{p_x^2+p_y^2}$. The angle brackets denote an
average over all considered particles from all events. The $v_1$ and
$v_2$ have complex multi-dimensional dependence since they are
functions of the $p_t$ and the normalized longitudinal rapidity
$y_0$ ($\equiv y_z/y_{pro}$ where
$y_z=\frac{1}{2}\ln\frac{E+p_z}{E-p_z}$ and the subscript $pro$ denotes the incident projectile in the center-of-mass system) from a colliding system with
certain beam energy $E_{lab}$ and impact parameter $b$ (or, the reduced one defined as $b_0=b/b_{max}$ with $b_{max} = 1.15 (A_{P}^{1/3} + A_{T}^{1/3})$ ). For mass-symmetric HICs,
the $v_1$ is an odd function of $y_0$. The variation of $v_1$ with
$y_0$ can be well described by a polynomial form $v_1(y_0)=\kappa
y_0 + by^3_0+c$, where $\kappa$ is the slope value at mid-rapidity
($y_0$=0) while $c$ is a constant which should be small enough to
ensure a good statistic average in flow calculations. In this work, more than 500,000 events of Au+Au collisions are simulated for each case.

 \begin{figure}[htbp]
 \centering
 \includegraphics[angle=0,width=0.9\textwidth]{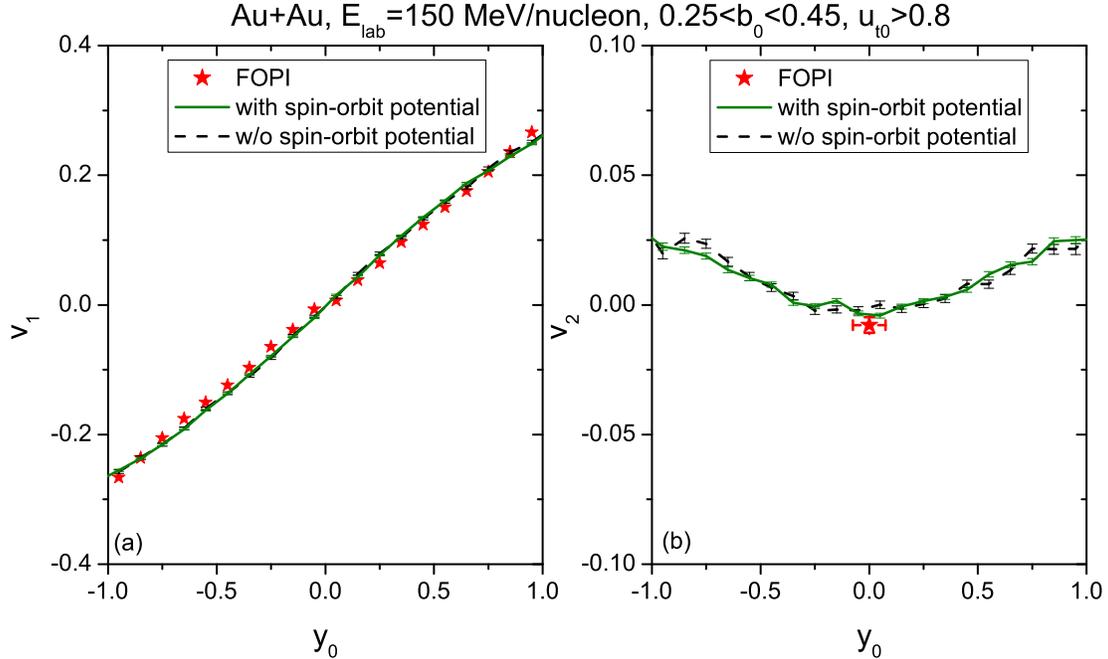}
 \caption{\label{fig1} (Color online) Directed flow $v_1$ (a) and elliptic flow $v_2$ (b) of free protons from Au+Au collisions
 at $E_{lab}$=150 MeV/nucleon and centrality $0.25<b_0<0.45$ as a function of the normalized rapidity $y_0$.
 Calculations with (solid lines) and without (dashed lines) spin-orbit potential are compared with the FOPI data (solid stars) reported in Ref.~\cite{FOPI:2011aa}.
 }
 \end{figure}

Fig.~\ref{fig1} shows the directed (a) and elliptic (b) flows of
free protons in semicentral ($0.25<b_0<0.45$) Au+Au collisions at 150 MeV$/$nucleon as simulated with (solid lines) and without (dashed lines) the spin-orbit potential.
The chosen
scaled cut $u_{t0}\equiv u_t/u_{pro}>0.8$ is the same as for the experimental data (solid stars) taken
from Ref.~\cite{FOPI:2011aa}.
Here $u_t=\beta_t\gamma$ is the transverse component of the four-velocity $u$=($\gamma$, $\bf{\beta}\gamma$).
It is clearly seen that simulations can reproduce
the FOPI data of both directed and elliptic flows quite well, especially for the directed flow in the whole rapidity region.
Further, it is noticed that both flows calculated with and without the spin-orbit term overlap almost
completely, which implies that the net contribution under a
spin-mixed circumstance is negligible to final flows. When the spin saturation is fulfilled in the same system of rectangular coordinates, the net contribution of spin-orbit term to emitted collective flows will be cancelled almost entirely by both the spin-up and spin-down nucleons.

\begin{figure}[htbp]
\centering
\includegraphics[angle=0,width=0.9\textwidth]{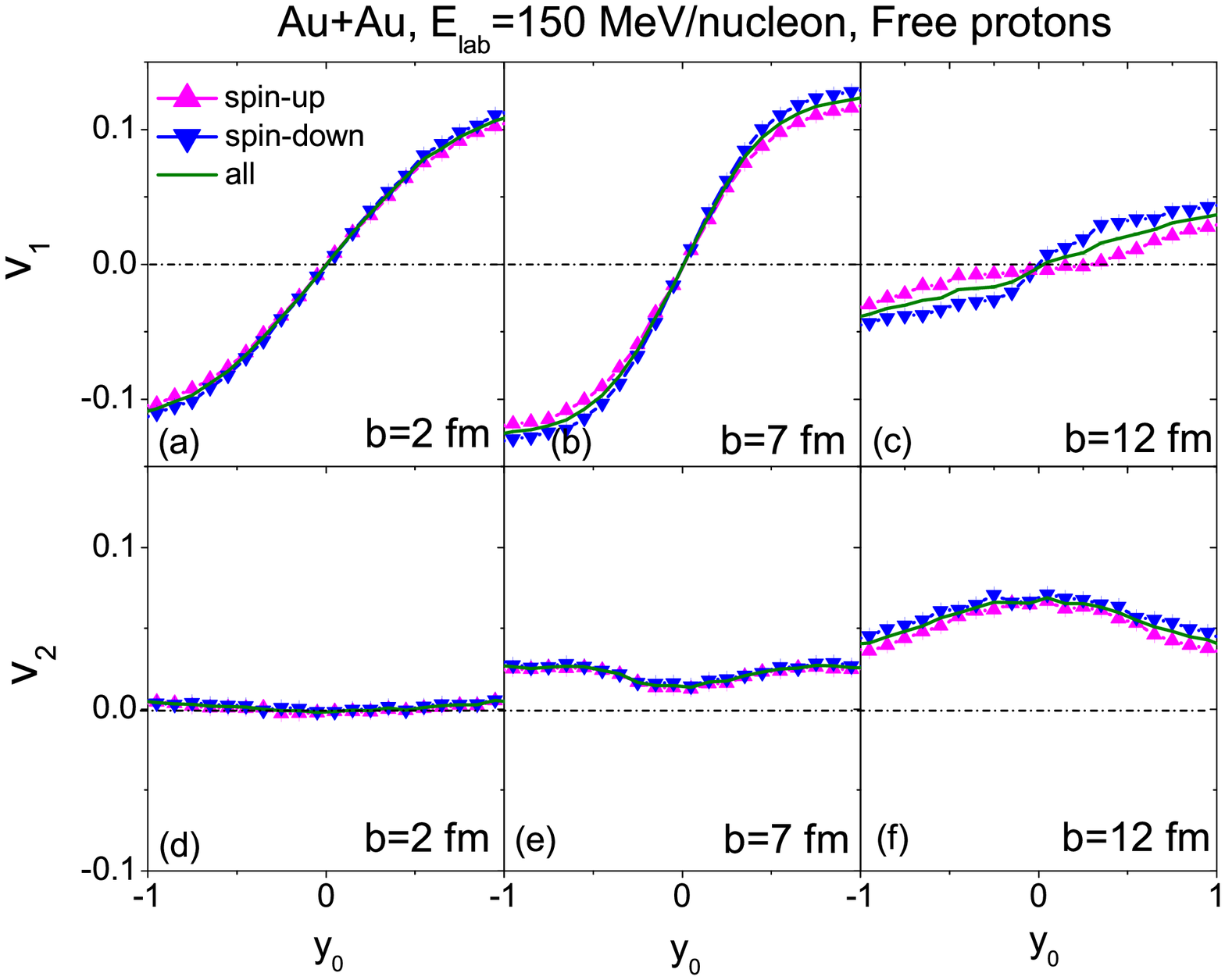}
\caption{\label{fig2}(Color online) Rapidity distribution of the
directed flow $v_1$ [upper panels (a), (b), (c)] and the elliptic
flow $v_2$ [lower panels (d), (e), (f)] of free protons from Au+Au
collisions at 150 MeV$/$nucleon for three centralities. Lines with up-triangles, solid lines, and lines with down-triangles represent flows of spin-up protons,
all free protons, and spin-down protons, respectively. }
\end{figure}

\begin{figure}[htbp]
\centering
\includegraphics[angle=0,width=0.9\textwidth]{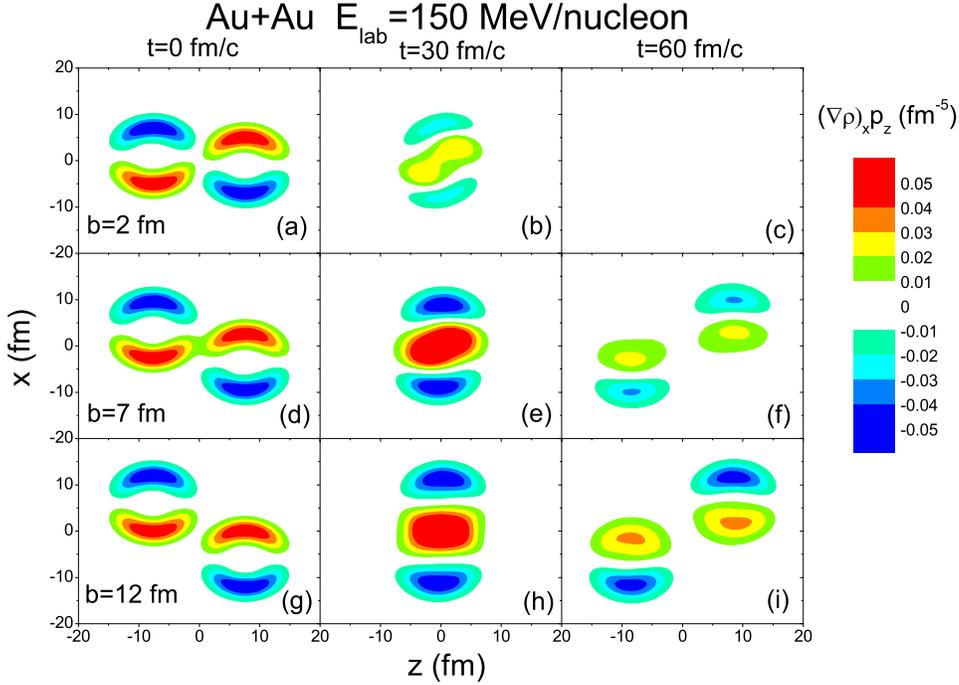}
\caption{\label{fig3} (Color online) Time evolution of $(\nabla\rho)_x$$p_z$ in the reaction $x-z$ plane produced from Au+Au collisions at 150 MeV$/$nucleon with impact parameters \emph{b}=2 fm [upper panels (a), (b), (c)], \emph{b}=7 fm [middle panels (d), (e), (f)], and \emph{b}=12 fm [lower panels (g), (h), (i)]. In [(c)] it is blank because the values of $(\nabla\rho)_x$$p_z$ lie between 0.01 and -0.01 $fm^{-5}$.
}
\end{figure}

In order to understand the net contribution of spin-orbit
potential to flows, in Fig.~\ref{fig2} we further show the directed
[upper panels (a),(b),(c)] and elliptic [lower panels (d),(e),(f)]
flows of both spin-up (lines with up-triangles) and spin-down (lines
with down-triangles) protons from Au+Au collisions at 150
MeV$/$nucleon and with several impact parameters \emph{b}=2, 7, and
12 fm, as well as flow results of spin-mixed ones (dubbed as ``all'', solid lines).
First, we see a flow splitting between
spin-up and spin-down protons, and flows of all free protons lie roughly between those of spin-up
and spin-down protons. Hence, because of the strong cancellation between spin-up and spin-down protons, the spin-orbit term shows almost no effect on  the flow of all free protons.
Second, the directed flow of spin-up protons is slightly smaller than that of spin-down protons, indicating that the spin-orbit potential
provides an additional attraction (repulsion) for
spin-up (spin-down) protons. The opposite sign of the spin-orbit potential for spin-up protons as compared to
spin-down protons helps to explain this phenomenon.
The elliptic flow does not show a
significant difference between the spin-up and spin-down protons, which is due to the large cancellation effect in the transverse momentum components as defined in Eq.~(\ref{eqv2}). And, it is known that the elliptic flow is strongly related to the nucleon-nucleon scattering and the blocking by the spectator matter, and
the dynamic evolution (densities and pressures achieved in the collision, and total collision number, etc.) does not change too much when
the spin-orbit term is taken into account.
Finally, as the impact parameter becomes larger, a larger flow splitting is observed. This centrality dependence of the attraction or repulsion  is in line with that shown in Ref.~\cite{xu:2014} and can be understood from the fact that the spin-orbit potential plays a more important role with larger impact parameters, which will be discussed in Fig.~\ref{fig3}.

Fig.~\ref{fig3} demonstrates the time evolution (shown with three time points: 0, 30, and 60 fm/\emph{c}) of the quantity $(\nabla\rho)_x$$p_z$ in the reaction plane for the same colliding system used in  Fig.~\ref{fig2}, which is the most important factor in determining the strength of the spin-orbit potential to flows. For a good demonstration effect, 1000 events of Au+Au collisions are used for each case.
It is clear that, in the time span from $t$=30 to 60 fm/\emph{c}, which is the main period of the flow formation, the value of $(\nabla\rho)_x$$p_z$ is positive (negative) in the center (outside) area. Since flows are formed mainly from the expansion of the central high-density region (followed by the neck fragmentation process), together with the consideration of the sign of spin-orbit interaction shown in Eqs.~(\ref{eq2}) and (\ref{eq3}), it comes to the conclusion that the spin-orbit term provides a net attraction (repulsion) for
spin-up (spin-down) protons. Further, due to a longer duration of the density gradient and a weaker nuclear stopping for collisions with a larger impact parameter, the $(\nabla\rho)_x$$p_z$ shows the largest effect on the flow from peripheral collisions (here, e.g., with $b=12$ fm).

\begin{figure}[htbp]
\centering
\includegraphics[angle=0,width=0.7\textwidth]{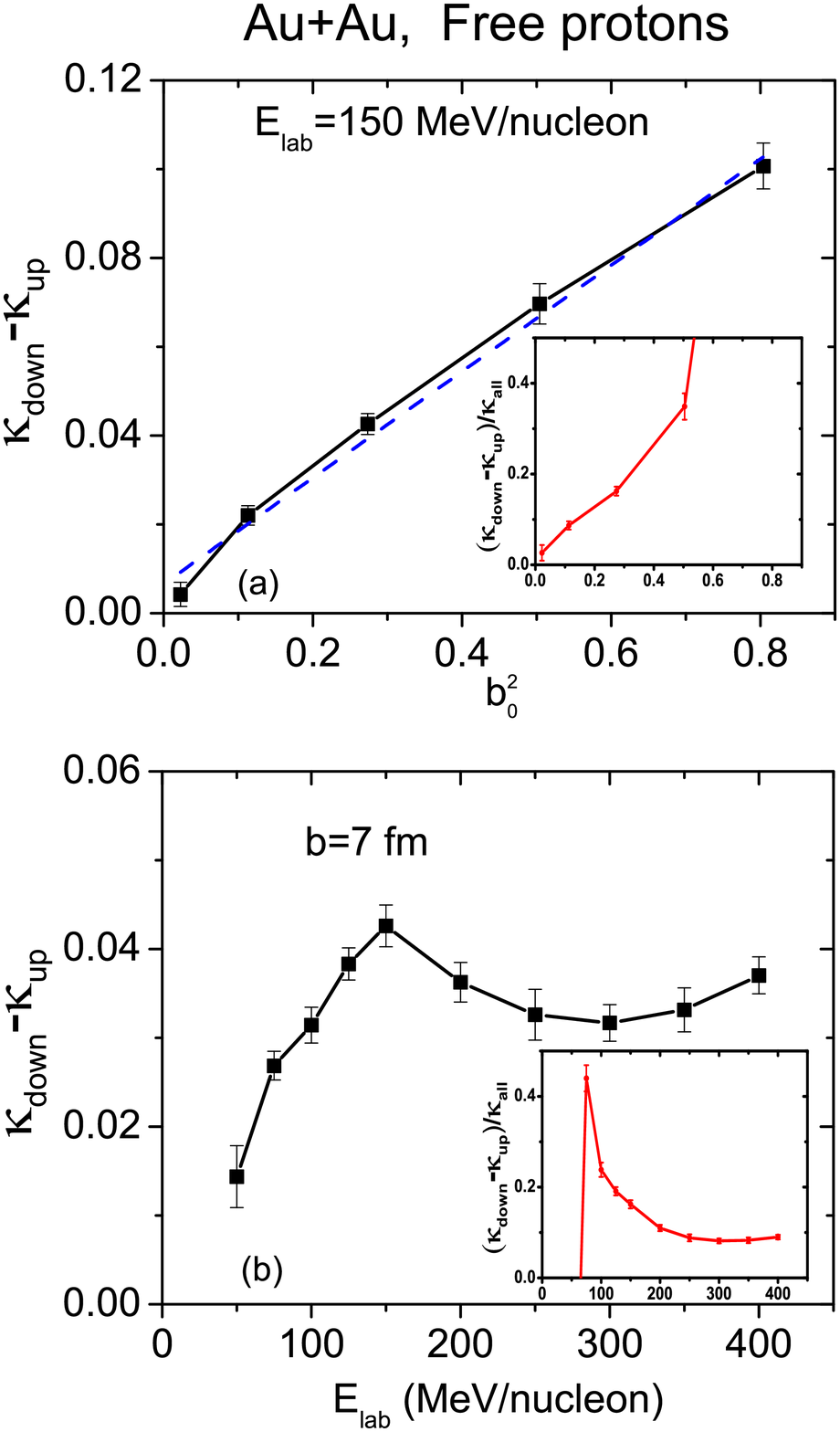}
\caption{\label{fig4} (Color online)
Centrality dependence (a) and excitation function (b) of the slope difference $\kappa_{down}$-$\kappa_{up}$ between spin-up and spin-down protons. The dashed line in (a) gives a linear fit to the calculation.
The inset in each panel shows the corresponding ($\kappa_{down}$-$\kappa_{up}$)/$\kappa_{all}$ ratio.}
\end{figure}

To exhibit more systematically the splitting of flows between spin-up and spin-down protons, Fig.~\ref{fig4} gives the slope difference $\kappa_{down}$-$\kappa_{up}$ from Au+Au collisions as a function of centrality $b_0^2$ at $E_{lab}$=150 MeV$/$nucleon (a) and of $E_{lab}$ at $\emph{b}=7 fm$ (b), respectively. In Fig.~\ref{fig4} (a), the $\kappa_{down}$-$\kappa_{up}$ is seen to increase almost linearly with increasing $b_0^2$, because the effect of spin-orbit term is controlled by the spatio-temporal range of the quantity $(\nabla\rho)_x$$p_z$ as illustrated in Fig.~\ref{fig3}. While in Fig.~\ref{fig4} (b), with the increase of beam energy, the $\kappa_{down}$-$\kappa_{up}$ first increases rapidly from 50 MeV$/$nucleon till to about 150 MeV$/$nucleon, then decreases with a following gentle plateau. As one expects, although the $(\nabla\rho)_x$$p_z$ certainly increases with increasing beam energy, the time duration becomes shorter. Moreover, the increased nucleon-nucleon scattering at higher beam energies weakens relatively the mean-field effect to some extent. The increase or decrease of the value of $\kappa_{down}$-$\kappa_{up}$ with increasing beam energy depends on the balance of these opposite influences. It is interesting to see that the peak at $\sim150$ MeV$/$nucleon is in agreement with the IBUU simulation in which the largest spin up-down differential transverse flow was observed at $E_{lab}$=100 MeV$/$nucleon for a similar collision ~\cite{xu:2014}.

The relative impact of the flow splitting $\kappa_{down}$-$\kappa_{up}$ can be seen more clearly from the ratio ($\kappa_{down}$-$\kappa_{up}$)/$\kappa_{all}$ versus $b_0^2$ and versus $E_{lab}$, which is shown in the inset of Fig.~\ref{fig4} (a) and (b), respectively. From the inset in Fig.~\ref{fig4} (a), one finds that the ratio increases more rapidly when $b_0^2\gtrsim0.5$ (or $b\gtrsim 9.5$ fm), which is due to the quick decrease of $\kappa_{all}$ at large impact parameters as shown in Fig.~\ref{fig2}. It is further found that the ratio reaches 0.1 and 0.3 when $b$ is larger than 4.5 fm and 9.5 fm, respectively. From the inset of Fig.~\ref{fig4} (b), one finds a sharp jump at $E_{lab}\sim$ 75 MeV$/$nucleon, which is due to the sign change (from negative to positive) and disappearance of the directed flow of protons \cite{Guo:2012aa} around this energy. It is further found that when $E_{lab}$ increases from 75 up to 200 MeV$/$nucleon, the ratio decreases but keeps still larger than 0.1. In view of the fact that the uncertainty in most sensitive probes of the density dependent symmetry energy is typically of the order of magnitude of 0.1, it is necessary to pay more attention to the influence of the spin-orbit interaction on the sensitivity of symmetry-energy related probes.

\begin{figure}[htbp]
\centering
\includegraphics[angle=0,width=0.9\textwidth]{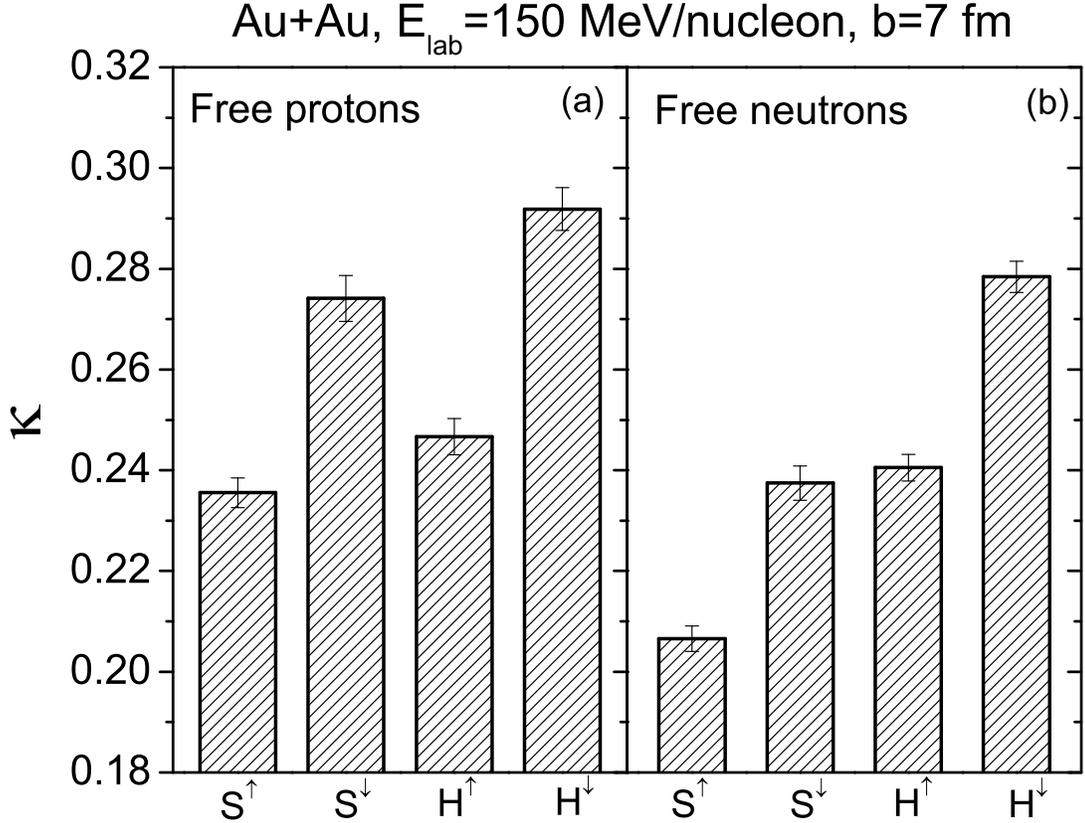}
\caption{\label{fig5} (Color online)
Slopes of the directed flow of spin-up (``$\uparrow$") and spin-down (``$\downarrow$") protons (a) and neutrons (b) calculated with the soft (``S", $\gamma$=0.5) and stiff (``H", $\gamma$=1.5) symmetry potential energies. }
\end{figure}

In Ref.~\cite{Guo:2012aa}, we found that the directed flow as well as its balance energy of neutrons from HICs is sensitive to the density dependence of symmetry potential energy while that of protons is not. In the current work, we have also found that the sensitivity of the directed flow of spin-polarized nucleons to the spin-orbit interaction is significant. So, in Fig.~\ref{fig5} we further compare slope values of the directed flow of spin-up and spin-down protons (a) and neutrons (b), as calculated with soft ($\gamma$=0.5) and stiff ($\gamma$=1.5) symmetry potential energies, where the symmetry potential energy is temporarily set to an exponential form $(\rho/\rho_0)^\gamma$.
It can be seen that the proton flow is obviously influenced ($>15\%$) by the spin orientation but weakly influenced ($<7\%$) by the stiffness of symmetry potential energy. While for the neutron flow case, both effects are on the same order, so that the $\kappa$ value of ``$S^\downarrow$" is almost the same as that of ``$H^\uparrow$". Meanwhile, the $\kappa$ value of ``$H^\downarrow$" is larger than that of ``$S^\uparrow$" by more than $33\%$. Therefore, the polarization of neutrons will obviously influence the sensitivity of neutron directed flow to the density dependence of symmetry energy. The spin-orbit interaction does not visibly influence spin-averaged experimental observables (as shown in Fig.~\ref{fig1}), but it will influence the determination of the density-dependent symmetry energy in heavy-ion reactions with the spin-polarized beam, which certainly deserves attention.

\section{Summary and Outlook}
In summary, with the help of the Skyrme potential energy density functional, the spin-orbit potential is introduced to the UrQMD model, and its effect on collective flows in Au+Au collisions at various beam energies and impact parameters have been studied. It is found that the spin-orbit potential generates a flow splitting (especially in the directed flow) between spin-up and spin-down nucleons, and the magnitude of the flow splitting depends on the impact parameter and incident energy. The directed flow of protons from HICs with beam energy around 150 MeV/nucleon and with a large impact parameter is suggested for future measurement to probe the in-medium spin-orbit interaction. We also find that the effect of spin-orbit potential is comparable to that of symmetry energy on the directed flow of neutrons.

Incorporating the spin-orbit term into the transport model is just the first step. Further investigations, such as the density and isospin dependence of the spin-orbit interaction and its influence in intermediate-energy HICs with various projectile-target combinations, are in progress.

\begin{acknowledgements}
We acknowledge support by the computing server C3S2 in Huzhou University.
This work was supported by the National Natural Science
Foundation of China under Grants Nos. 11025524, 11375062 and 11161130520, National Basic Research Program of China
under Grant No. 2010CB832903, the European Commission¡¯s 7th Framework Programme (FP7-PEOPLE-2010-IRSES) under Grant Agreement Project No. 269131,  and the project sponsored by SRF for ROCS, SEM.

\end{acknowledgements}

\end{document}